\RequirePackage{lineno}
\documentclass[twocolumn,10pt,5p]{elsarticle}
\biboptions{sort&compress}
\usepackage{subfigure}
\usepackage{epsfig}
\usepackage{dcolumn}
\usepackage{bm}
\usepackage{graphicx}
\usepackage{endnotes}
\usepackage{multirow}
\usepackage{xspace}
\usepackage{atlasphysics}
\RequirePackage{hypernat}
\RequirePackage[colorlinks,breaklinks,pdftitle={ATLAS draft},pdfauthor={The ATLAS Collaboration}]{hyperref}  
\hypersetup{linkcolor=blue,citecolor=blue,filecolor=black,urlcolor=blue}

\newcommand{\myTitle}{Search for Scalar Top Quark Pair Production in Natural Gauge Mediated Supersymmetry Models with the ATLAS Detector \\ in $pp$ Collisions at $\sqrt{s}=$~7~TeV}

\usepackage{preprintcover}
\PreprintCoverPaperTitle{\myTitle}
\PreprintIdNumber{CERN-PH-EP-2012-097}
\PreprintCoverAbstract{
The results of a search for pair production of the lighter scalar partners of top quarks ($\tone$) 
in \totallumi~of $pp$ collisions at $\sqrt{s}=7$~TeV using the ATLAS experiment at the LHC are 
reported. Scalar top quarks are searched for in events with two same flavour opposite-sign 
leptons ($e,\mu$) with invariant mass consistent with the $Z$ boson mass, large missing 
transverse momentum and jets in the final state. At least one of the jets is identified 
as originating from a $b$-quark. No excess over Standard Model expectations is found. 
The results are interpreted in the framework of $R$-parity conserving, gauge mediated 
Supersymmetry breaking `natural' scenarios, where the neutralino ($\tilde{\chi}_{1}^{0}$) is the 
next-to-lightest supersymmetric particle. Scalar top quark masses up to 310~GeV are excluded for 
115~GeV~$< m_{\tilde{\chi}_{1}^{0}}<$~230~GeV at 95\% confidence level, reaching an 
exclusion of $m_{\tone}<$~330~GeV for $m_{\tilde{\chi}_{1}^{0}}=190$~GeV. 
Scalar top quark masses below $240$~GeV are excluded for all values of $m_{\tilde{\chi}_{1}^{0}}>m_{Z}$.
}
\PreprintJournalName{Physics Letters B}

\newcommand{\sqR}   {\tilde{q}^{}_{R}}
\newcommand{\sqL}   {\tilde{q}^{}_{L}}

\newcommand{\tone}   {\tilde{t}^{}_{1}}

\newcommand{\neut}  {\tilde{\chi}^{0}_{1}}  
  
\newcommand{\chipm}  {\tilde{\chi}^{\pm}_{1}} 
\newcommand{\gravitino}{\ensuremath{\tilde{G}}\xspace}
\def\etmiss{\ensuremath{E_{\mathrm{T}}^{\mathrm{miss}}}\xspace}

\def\bjets{{\it b-}jets\xspace}

\def\btagging{{\it b-}tagging\xspace}

\def\bquarks{{\it b-}quarks\xspace}

\def\Wjets{$W$+jets\xspace}

\def\ZHF{$Z$+hf\xspace}
\renewcommand{\ttbar} {\ensuremath{t\bar{t}}\xspace}
\newcommand{\ttbb}{\ensuremath{t\bar{t}b\bar{b}}\xspace}
\renewcommand{\met} {\ensuremath{E_{\mathrm{T}}^{\mathrm{miss}}}\xspace}
\renewcommand{\pt} {\ensuremath{p_\mathrm{T}}\xspace}

\newcommand{\lumi}{2.05}
\newcommand\totallumi{2.05~${\rm fb}^{-1}$}
\def\MET{\ensuremath{E_{\mathrm{T}}^{\mathrm{miss}}}\ }
\def\met{\ensuremath{E_{\mathrm{T}}^{\mathrm{miss}}}}

\begin{document}
\title{\myTitle}
\author{The ATLAS Collaboration}

\journal{Physics Letters B}

\begin{abstract}
The results of a search for pair production of the lighter scalar partners of top quarks ($\tone$) 
in \totallumi~of $pp$ collisions at $\sqrt{s}=7$~TeV using the ATLAS experiment at the LHC are 
reported. Scalar top quarks are searched for in events with two same flavour opposite-sign 
leptons ($e,\mu$) with invariant mass consistent with the $Z$ boson mass, large missing 
transverse momentum and jets in the final state. At least one of the jets is identified 
as originating from a $b$-quark. No excess over Standard Model expectations is found. 
The results are interpreted in the framework of $R$-parity conserving, gauge mediated 
Supersymmetry breaking `natural' scenarios, where the neutralino ($\tilde{\chi}_{1}^{0}$) is the 
next-to-lightest supersymmetric particle. Scalar top quark masses up to 310~GeV are excluded for 
115~GeV~$< m_{\tilde{\chi}_{1}^{0}}<$~230~GeV at 95\% confidence level, reaching an 
exclusion of $m_{\tone}<$~330~GeV for $m_{\tilde{\chi}_{1}^{0}}=190$~GeV. 
Scalar top quark masses below $240$~GeV are excluded for all values of $m_{\tilde{\chi}_{1}^{0}}>m_{Z}$.

\end{abstract}

\maketitle
\normalsize

\section{Introduction}
Supersymmetry (SUSY)~\cite{Miyazawa:1966,Ramond:1971gb,Golfand:1971iw,Neveu:1971rx,Neveu:1971iv,Gervais:1971ji,Volkov:1973ix,Wess:1973kz,Wess:1974tw}
provides an extension to the Standard Model (SM) which can resolve the hierarchy problem. 
For each known boson or fermion, SUSY introduces a particle (sparticle) with identical quantum 
numbers except for a difference of half a unit of spin. 
The non-observation of the sparticles implies that SUSY is broken and the  
superpartners are generally heavier than the SM partners. In the framework of a 
generic $R$-parity conserving minimal supersymmetric extension of the SM (MSSM)~\cite{Fayet:1976et,Fayet:1977yc,Farrar:1978xj,Fayet:1979sa,Dimopoulos:1981zb}, 
SUSY particles are produced in pairs and the lightest supersymmetric particle (LSP) is stable. 

The scalar partners of right-handed and left-handed quarks, $\sqR$ and $\sqL$, 
can mix to form two mass eigenstates. In the case of the scalar top quark ($\tilde{t}$, stop), 
large mixing effects due to the Yukawa coupling, $y_t$, and the trilinear coupling, $A_{t}$, 
can lead to one stop mass eigenstate, $\tone$, that is significantly lighter than 
other squarks. Consequently, the $\tone$ could be produced with large cross sections 
at the LHC via direct pair production. 

Light stop masses are favoured by arguments of `naturalness' of electroweak 
symmetry breaking~\cite{Kitano:2006}, because of the possibly large coupling between 
the $\tilde{t}$ and the Higgs boson, $h$. In particular, radiative corrections to the Higgs boson mass  
mainly arise from the stop-top loop diagrams including top Yukawa and three-point stop-stop-Higgs interactions.

In gauge mediated SUSY breaking~(GMSB) models~\cite{Dine:1981gu,AlvarezGaume:1981wy,Nappi:1982hm,Dine:1993yw,Dine:1994vc,Dine:1995ag},   
 gauge interactions (messengers) are responsible for the appearance of soft 
supersymmetry breaking terms. If the characteristic scale of the masses of the 
messenger fields 
is about 10~\TeV, an upper bound on $m_{\tilde{t}_1}$ of about 
400~\GeV~is found when imposing the absence of significant ($\sim 10$\%) fine tuning~\cite{Kitano:2006}.  

In GMSB, the gravitino \gravitino is the LSP (in general $m_{\gravitino}\muchless 1$~\keV). 
The experimental signatures are largely determined by the nature of the next-to-lightest SUSY 
particle~(NLSP). For several GMSB models the NLSP is the lightest neutralino, $\neut$,  
promptly decaying to its lighter SM partner through gravitino emission. 
Neutralinos are mixtures of gaugino ($\tilde{B}$, $\tilde{W}^{0}$) and higgsino 
($\tilde{H}^{0}_{u}$, $\tilde{H}^{0}_{d}$) gauge-eigenstates, and therefore the lightest 
neutralino decays to either a $\gamma$, $Z$ or Higgs boson. 
If the $\neut$ is higgsino-like, it decays either via $\neut \rightarrow h \gravitino$ or 
$\neut \rightarrow Z \gravitino$.  Light higgsinos 
lead to a large higgsino component in $\neut$ 
and a small mass difference between $\neut$ and $\chipm$. In particular, if the higgsino mass 
($|\mu|$) is much smaller than the gaugino masses (pure higgsino case), $\neut$ and 
$\chipm$ are almost degenerate such that the ($ff'$) system resulting from  
the chargino decay $\chipm \rightarrow \neut ff'$ is very soft.  \\

In this Letter, a search for direct stop pair production is presented, 
assuming a GMSB model where the $\neut$ is purely 
higgsino-like 
and is lighter than the $\stop_{1}$~\cite{Asano:2010}. The model parameters are 
\begin{equation}
m_{\tilde{q}_3} = m_{\tilde{u}_3} = - A_{t}/2;  \, \, \, \, \, \, \rm{tan \beta} = 10,   
\end{equation}
where $m_{\tilde{q}_3}$ and $m_{\tilde{u}_3}$ are the soft SUSY breaking masses for the left- 
and right-handed third-generation squarks, respectively, and tan$\beta$ is the ratio of the vacuum 
expectation values of up-type and down-type Higgs field. 
In these scenarios, masses of first and second generation squarks and gluinos 
(superpartners of the gluons) are above 2 TeV,  the $\tilde{t}$ mass eigenstates 
are such that $m_{\tilde{t}_2}~\gg~m_{\tilde{t}_1}$ and only $\tone$ pair production 
is considered in what follows. 
Stops decay either via $\tilde{t}_1\rightarrow b\tilde{\chi}_{1}^{+}$ or, 
if kinematically allowed, via $\tilde{t}_1\rightarrow t\tilde{\chi}_{1(2)}^{0}$. 
For the scenarios considered, the subsequent decay $\neut \rightarrow Z \gravitino$ 
has a branching ratio (BR) between 1 and 0.65 for $m_{\neut}$ between 100~\GeV\, and 350~\GeV~\cite{Dimopoulos:1997}.  
Thus, the expected signal is characterised by the presence of two jets originating 
from the hadronisation of the \bquarks ({\it b-}jets), decay products of $Z$ (or $h$) 
bosons and large missing transverse momentum --- its magnitude is here referred 
to as $\etmiss$ --- resulting from the undetected gravitinos. 

This search uses data recorded between March and August 2011 by the ATLAS detector at the LHC. 
After the application of beam, detector, and data quality requirements, the 
dataset corresponds to a total integrated luminosity of 
$2.05\pm0.08$~fb$^{-1}$~\cite{Aad:2011dr,ATLAS:Lumi}. 
To enhance the sensitivity to the aforementioned SUSY scenarios, events are 
required to contain energetic jets, of which one must be identified as a {\it b-}jet, 
large $\etmiss$ and two opposite-sign, same flavour leptons ($\ell = e, \mu$) with 
invariant mass consistent with the $Z$ boson mass, $m_{Z}$. 
This is the first search for scalar top quarks decaying via $Z$ bosons in GMSB models.   
General searches for supersymmetric particles in events with a $Z$ boson, energetic 
jets and missing transverse momentum have been reported by the CMS 
collaboration~\cite{CMSZJetMet}. Searches for direct stop pair production have been 
performed at the CDF and D0 experiments assuming different 
SUSY mass spectra and decay modes (see for example Ref.~\cite{Aaltonen:2010st} and ~\cite{D0Stop}). 
 

\section{The ATLAS Detector}
The ATLAS detector~\cite{DetectorPaper:2008} consists of inner tracking devices
surrounded by a superconducting solenoid, electromagnetic
and hadronic calorimeters, and a muon spectrometer
with a toroidal magnetic field. 

The inner detector system, in combination with the 2~T field from 
the solenoid, provides precision tracking of charged particles for 
$|\eta| < 2.5$~\footnote{ATLAS uses a right-handed coordinate system with its origin at the nominal
  interaction point (IP) in the centre of the detector and the $z$-axis 
  coinciding with the axis of the beam pipe. The $x$-axis points from the IP to
  the centre of the LHC ring, and the $y$-axis points upward. Cylindrical
  coordinates $(r,\phi)$ are used in the transverse plane, $\phi$ being the
  azimuthal angle around the beam pipe. The pseudorapidity is defined in terms
  of the polar angle $\theta$ as $\eta=-\ln\tan(\theta/2)$.
  The distance~$\Delta R$ in the $\eta-\phi$ space is defined as
  $\Delta R = \sqrt{(\Delta\eta)^2+(\Delta\phi)^2}$.}.
It consists of a silicon pixel detector, a silicon microstrip detector  
and a straw tube tracker that also provides transition
radiation measurements for electron identification.
The calorimeter system covers the pseudorapidity range
$|\eta| < 4.9$. It is composed of sampling calorimeters with
either liquid argon (LAr) or scintillating tiles as the active
media.
The muon spectrometer surrounds the calorimeters. It consists of a set of 
high-precision tracking chambers placed within a magnetic field generated 
by three large superconducting eight-coil toroids. The spectrometer, which 
has separate trigger chambers for $|\eta| < 2.4$, provides muon identification 
and measurement for $|\eta| < 2.7$.

\section{Simulated Event Samples}\label{sec:SimEvSamp}
Simulated event samples are used to aid in the description of the background, as well as to 
determine the detector acceptance, the reconstruction efficiencies and the 
expected event yields for the SUSY signal. 

The signal samples are simulated  with the {\tt HERWIG++}~\cite{Bahr:2008pv} 
v2.4.2 Monte Carlo (MC) program at fixed $\tone$ and $\neut$ masses, obtaining the 
desired values by varying the $m_{\tilde{q}_3}$ and $|\mu|$ parameters. 
The particle mass spectra and decay modes are determined using {\tt ISASUSY} 
from the {\tt ISAJET}~\cite{isajet} v7.80 program. 
The SUSY sample yields are normalised to the results of next-to-leading-order (NLO) calculations 
as obtained using {\tt PROSPINO}~\cite{Beenakker:1997ut} v2.1 including higher-order supersymmetric 
QCD corrections and the resummation of soft-gluon emission at 
next-to-leading-logarithmic (NLL) accuracy~\cite{Beenakker:2010nq}.  
An envelope of cross-section predictions is defined using the 68\% C.L. ranges 
of the {\tt CTEQ6.6M}~\cite{cteq6m} (including $\alpha_s$ uncertainty) and 
{\tt MSTW2008}~\cite{Martin:2009iq} parton distribution function (PDF) sets, 
together with variations of the factorisation and renormalisation 
scales, set to the stop mass. The nominal cross section is taken to be the midpoint 
of the envelope and the uncertainty assigned is half the full width of the envelope, following 
closely the {\tt PDF4LHC} recommendations~\cite{PDF4LHC}.
NLO+NLL cross sections vary between 80~pb and 0.1~pb for stop masses 
between 140~\GeV\, and 450~\GeV.

For the backgrounds the following SM processes are considered. 
Top quark pair and single top quark production are simulated with {\tt MC@NLO}~\cite{mcatnlo}, 
setting the top quark mass to 172.5~\GeV, and using the NLO PDF set {\tt CTEQ6.6}~\cite{cteq66}.
Additional samples generated with {\tt POWHEG}~\cite{powheg} and {\tt ACERMC}~\cite{Kersevan:2004yg} 
are used to estimate the event generator systematic uncertainties.
Samples of \Wjets, $Z/\gamma^{*}+$jets with light- and heavy-flavour jets, and \ttbar with additional \bjets, \ttbb, 
are generated with {\tt ALPGEN}~\cite{alpgen} and the PDF set {\tt CTEQ6L1}~\cite{cteq6l1}.
The fragmentation and hadronisation for the {\tt ALPGEN} and {\tt MC@NLO} samples are
performed with {\tt HERWIG}~\cite{herwig}, using {\tt JIMMY}~\cite{jimmy} for the
underlying event.
Samples of $Z$\ttbar and $W$\ttbar are generated with {\tt MADGRAPH}~\cite{madgraph}
interfaced to {\tt PYTHIA}~\cite{pythia}. Diboson ($WW$, $WZ$, $ZZ$) samples
are generated with {\tt HERWIG}.
For the comparison to data, all SM background cross sections are normalised to the 
results of higher-order calculations using the same values as Ref.~\cite{PRD_gluinomed}.  

The MC samples are produced using {\tt PYTHIA} and {\tt HERWIG}/{\tt JIMMY} parameters tuned as described 
in Ref.~\cite{atlastune2010}  
and are processed through a detector simulation~\cite{atlassimulation} based on {\tt GEANT4}~\cite{geant4}. 
Effects of multiple proton--proton interactions~\cite{atlastune2010} are included in the simulation and MC events are 
re-weighted to reproduce the mean number of collisions per bunch crossing estimated from data.

\section{Object Reconstruction} 

Jet candidates are reconstructed using the anti-$k_t$ jet clustering 
algorithm~\cite{Cacciari:2006,Cacciari:2008gp} with a radius parameter of $0.4$. 
The inputs to the algorithm are three-dimensional calorimeter 
energy clusters~\cite{topoclusters} seeded by cells with energy calibrated at the 
electromagnetic energy scale significantly above the measured noise. 
The jet energy is corrected for inhomogeneities and 
for the non-compensating nature of the calorimeter using $p_{\rm T}$- and $\eta$-dependent 
correction factors derived using simulated multi-jet events (following Ref.~\cite{ATLAS:JES2011} and 
references therein). Only jet candidates with $\pt > 20$~\GeV{} and $|\eta| < 2.8$ are retained.

A {\it b-}tagging algorithm~\cite{ATLAS:BtagCalib} 
is used to identify jets containing a {\it b-}hadron decay. 
The algorithm is based on a multivariate technique based on properties of the secondary 
vertex, of tracks within the jet and of the jet itself.  
The nominal {\it b-}tagging efficiency, computed on $t\bar{t}$ MC events, is on average 60\%, with a 
misidentification (mis-tag) rate for light-quark/gluon jets of less than 1\%.  
These {\it b-}jets are identified within the nominal acceptance of the inner detector 
($|\eta| < 2.5$) and are required to have $\pt > 50$~\GeV{}. 

Electron candidates are required to have $\pt > 20$~\GeV{} and $|\eta| <$ 2.47, and 
are selected to satisfy the `tight' shower shape and track selection criteria 
of Ref.~\cite{ATLAS:electron_performance}.   
The candidate electron must be isolated, such that the \pt sum of tracks ($\Sigma \pt$, not including 
the electron track),  within a cone in the $(\eta, \phi)$ plane of radius $\Delta R = 0.2$ around the candidate 
must be less than 10\% of the electron \pt. 

Muons are reconstructed using an algorithm~\cite{ATLAS:muonreco:2010} 
which combines the inner detector and the muon spectrometer information (combined muons).
A muon is selected for the analysis only if it has $\pt>10$~GeV and $|\eta|<2.4$, and 
 the sum of the transverse momenta of tracks within a cone of $\Delta R=0.2$
around it is less than $1.8 $~GeV. 
To reject cosmic rays, muons are required to have longitudinal and transverse impact parameters within 1\,mm and 
0.2\,mm of the primary vertex, respectively.

Following the object reconstruction described above, overlaps between jet candidates and leptons 
are resolved. 
Any jet within a distance $\Delta R $ = 0.2 of a candidate electron is discarded. Any remaining 
lepton within $\Delta R=$ 0.4 of a jet is discarded. 

The $\etmiss$ is calculated from the vectorial sum of the transverse momenta of 
jets (with $p_{\rm T}>20$~GeV and $|\eta|<4.5$), electrons and muons --- including non-isolated 
muons~\cite{ATLAS:met_performance}. The four-vectors of calorimeter clusters 
not belonging to other reconstructed objects are also included.

During 40\% of the data-taking period, a localised electronics failure in the LAr barrel calorimeter 
created a dead region in the second and third calorimeter layers 
($\Delta\eta\times\Delta\phi \simeq 1.4\times 0.2$) in which, on average, 30\% of the incident 
energy is not measured. 
If a jet with $p_{\rm T}>50$~GeV or an electron candidate falls in this region, the event is rejected.
The loss in signal acceptance is less than 10\% for the models considered.

\section{Event Selection} 

The data are selected with a three-level trigger system based on the presence of leptons. Two trigger 
paths are considered: a single electron trigger, reaching a plateau efficiency for electrons with 
$\pt \geq 25$ GeV, and a combined muon+jet trigger, reaching a plateau efficiency for muons with 
$\pt \geq  20$ GeV and jets with $\pt \geq 60$ GeV. 

Events must pass basic quality criteria against detector noise and non-collision 
backgrounds~\cite{ATLAS:JES2011} are required to have a reconstructed primary vertex 
associated with five or more tracks; when more than one such vertex is found, the 
vertex with the largest summed $p^{2}_{\rm T}$ of the associated tracks is chosen

The selections applied in this analysis are listed below.
\begin{itemize}
\item To ensure full efficiency of the trigger, events are selected if they contain at least one electron 
      with $p_{\rm T}>25$~GeV or one muon with $p_{\rm T}>20$~GeV. 
         
\item Exactly two same flavour opposite-sign leptons ($ee,\mu\mu$) are required, such that their  
      invariant mass $m_{\ell\ell}$ is within the $Z$ mass range (86 GeV $<m_{\ell\ell}<$ 96 GeV). 
      Events with additional electron or muon candidates are vetoed.  

\item Events must include at least one jet with $p_{\rm T}>60$~GeV and one additional jet with $p_{\rm T}>50$~GeV.

\item At least one jet with $p_{\rm T}>50$~GeV and $|\eta|<2.5$  is required to be {\it b-}tagged.
\end{itemize}
                  
Two signal regions, referred to as SR1 and SR2, are defined 
using two different $\etmiss$ threshold requirements in order to maximise the  
sensitivity across the $\tone$-$\neut$ mass plane. For SR1, $\etmiss > 50$~GeV is required and it is chosen  
for models with $\Delta m = m_{\tone}-m_{\neut}$ larger than 100~GeV or $m_{\tone}<300$~GeV, where moderate missing 
transverse momentum is expected. SR2 is optimised for small $\Delta m$ scenarios and events are required 
to have $\etmiss > 80$~GeV. 
 
The signal efficiencies, which include the $Z\rightarrow ee,\mu\mu$ BR, acceptance and detector effects, vary 
across the $\tone$-$\neut$ mass plane. For SR1 (SR2) the efficiencies are found to lie 
between 0.03\% and 2.1\% (0.01\% and 1.7\%) as the stop mass increases from 
140~GeV to 400~GeV, and between 0.6\% and 2.0\% (0.5\% and 1.7\%) for $\Delta m$ between 300~GeV and 100~GeV   
at a stop mass of 400~GeV.

\section{Background Estimation} 
The main SM processes contributing to the background are, in order of importance, top quark pair 
and single top quark production, followed by the associated production of $Z$ bosons and 
heavy-flavour jets --- referred to as \ZHF. 

\begin{figure}[t!]
  \begin{center}
    \includegraphics[width=8cm]{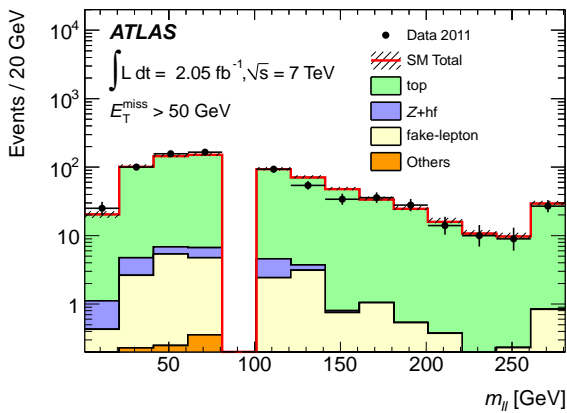}
      \end{center}
  \caption{The distribution of $m_{\ell\ell}$ in CR1 for the sum of $ee$ and $\mu\mu$ channels. 
  The dashed band shows the experimental systematic uncertainties including effects due to JES, 
  \btagging and lepton ID efficiency. The last $m_{\ell\ell}$ bin includes the number of overflow 
  events for both data and SM expectation. }
  \label{figure:topCR_plots}
\end{figure}

The top background is evaluated using control regions (CRs) that are the same as the SRs with the 
exception of the $m_{\ell\ell}$ requirement (modified to 15~GeV $<m_{\ell\ell}<$ 81~GeV or $m_{\ell\ell}>$ 101~GeV). 
Depending on the corresponding signal region, CRs are labelled as CR1 and CR2. 
In both cases, negligible yields from the targeted SUSY signals are expected. 
The background estimation in each SR is obtained by multiplying the number of events observed 
in the corresponding CR --- corrected using simulations for non-top backgrounds --- by a transfer 
factor, defined as the ratio of the MC-predicted yield in the signal region to that in the control region:
\begin{equation}
  N_{\rm SR}^{\rm top} = \frac{ N^{\rm top,MC}_{\rm SR} }{ N^{\rm top,MC}_{\rm CR} }
    \left( N_{\rm CR}^{\rm obs} - N_{\rm CR}^{\rm non\mathchar`-top,MC} \right)
\label{eq2}
\end{equation}
where $N_{\rm CR}^{\rm obs}$ denotes the observed yield in the CR. For each CR, the contribution from 
other SM processes accounts for less than 10\% of the total.
The estimate based on this approach benefits from a cancellation of systematic uncertainties that are 
correlated between SRs and CRs. 
The distribution of $m_{\ell\ell}$ for CR1 is shown in Figure~\ref{figure:topCR_plots}. 
The experimental uncertainties, described in Section~\ref{sec:syst}, are displayed. They include 
effects due to jet energy scale and resolution~\cite{ATLAS:JES2011} (JES),
\btagging~\cite{ATLAS:BtagCalib} and lepton identification (ID) efficiencies~\cite{ATLAS:electron_performance,ATLAS:muonreco,ATLAS:muonreco:2010}. 
The number of expected events for 2.05~fb$^{-1}$ of integrated 
luminosity as predicted by the MC simulation is in good agreement with data for
both CRs without introducing data/MC scaling factors.

\begin{figure}[!t]
  \centering	  
  \includegraphics[width=8cm]{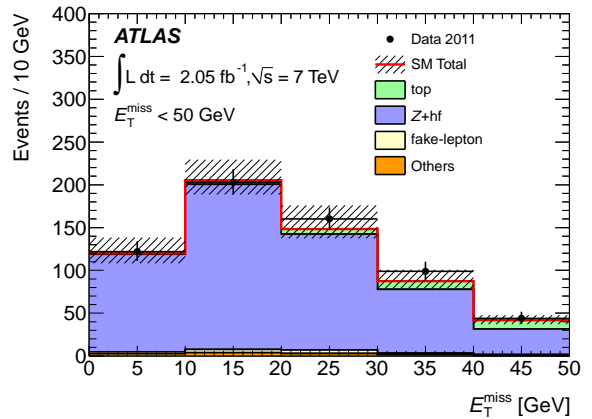}
\caption{The distribution of $\etmiss$ for the \ZHF~validation region for the sum of $ee$ and $\mu\mu$ channels. 
         By construction, only events with $E_{\rm T}^{\rm miss}$~below 50~GeV are displayed. 
         The dashed band represents the experimental uncertainties
	 including effects due to JES, \btagging and lepton ID efficiency.}
\label{figure:ZCR_plots}
\end{figure}

The topology of \ZHF~production events is similar to that of the signal, especially in low $\tone$-$\neut$ 
mass scenarios. Therefore the background from the \ZHF~process is estimated from MC simulation 
and validated in a control region where events passing all SR selection criteria except for a 
reversed \met~cut ($E_{\rm T}^{\rm miss}<50$~GeV) are considered. 
Possible signal contamination in the control region varies across the $\tone$-$\neut$ mass range. 
As an example, for $m_{\tilde{\chi}_{1}^{0}} \simeq 100$~GeV, the contamination is 5\% (80\%) of 
the total predicted SM background for $m_{\tone} \simeq 350$ (150)~GeV. 
In Figure~\ref{figure:ZCR_plots} the $E_{\rm T}^{\rm miss}$ distribution is shown in the range 0--50~GeV 
for $ee$+$\mu\mu$ final states. The number of events observed in data is in good agreement 
with the SM expected yields within experimental uncertainties.    

Backgrounds from {\it W}+jets and multi-jet production, referred to as ``fake-lepton'' 
contributions, are subdominant. In this case, events passing the selection contain at 
least one misidentified or non-isolated lepton (collectively called ``fakes"). The fake-lepton background 
estimate is obtained using the data-driven approach described in Ref.~\cite{toppaper}.  
The probability of misidentifying a jet as a signal lepton is estimated in control 
regions dominated by multi-jet events where exactly one pre-selected lepton, at least 
one {\it b-}tagged jet and low $E_{\rm T}^{\rm miss}$ are required.  

Finally, background contributions from diboson, $Z$\ttbar, $W$\ttbar and \ttbb~events 
--- referred to as `Others' --- are estimated from MC simulation. They account for less 
than 3\% of the total SM background in either SR.

\section{Systematic Uncertainties}\label{sec:syst}
Various systematic uncertainties affecting the background rates and 
signal yields have been considered. The values quoted in the following 
refer to $ee$ and $\mu\mu$ channels summed.   

Systematic uncertainties on the top background expectations vary between 11\% and 13\% depending 
on the SR and are dominated by the residual uncertainties on the shape of the  
kinematic distributions of top quark events. The uncertainties are evaluated using additional MC samples. 
{\tt ACERMC}~\cite{Kersevan:2004yg} is used to evaluate the impact of initial and 
final state radiation parameters (varied as in Ref.~\cite{topcross_dilep}), {\tt PYTHIA} 
for the choice of fragmentation model, {\tt POWHEG}~\cite{powheg} for the choice of generator. 
Experimental uncertainties on the \btagging~efficiency, JES and lepton ID 
efficiency account for about 4\% in either SR. 

The dominant uncertainties on the \ZHF~background estimates from simulation 
arise from the uncertainty on the production cross section used to 
normalise the MC yields. A $\pm 55$\% uncertainty on the total production cross 
section is evaluated from the direct \ZHF~inclusive measurement described in 
Ref.~\cite{ATLAS:Z+bjet} and takes into account differences between 
data, {\tt MCFM}~\cite{Campbell:2004} and {\tt ALPGEN} predictions. 
The extrapolation to each following jet multiplicity in $Z+hf+$N jets events 
increases this uncertainty by an additional 24\%~\cite{Alwall:2008}. 
Other uncertainties due to JES, {\it b-}tagging efficiency and lepton ID efficiency  are 
found to be about 25\% and 35\% for SR1 and SR2, respectively. 

The estimated fake-lepton background is affected by systematic uncertainties 
related to the determination of the lepton misidentification rate and to the
subtraction of non-multi-jet contributions to the event yield in the multi-jet 
enhanced regions. The estimated uncertainty is 50\% and 60\% in SR1 and SR2, 
respectively. 
Finally, a conservative 100\% uncertainty is taken into account on the 
contributions from `Others'. 

For the SUSY signal processes, uncertainties on the renormalisation and 
factorisation scales, on the PDF and on $\alpha_{s}$ affect the cross section 
predictions. PDF and $\alpha_{s}$ uncertainties
are between 10\% and 15\% depending on $m_{\tone}$ for the mass range 
considered.  The variation of renormalisation and factorisation scales by 
a factor of two changes the nominal signal cross section by 9--13\%
depending on the stop mass.
The impact of detector-related uncertainties, such as JES, \btagging and lepton ID 
efficiency, on the signal event yields varies between 10\% and 25\% 
and is dominated by the uncertainties on the JES.

\section{Results and Interpretation}
The numbers of observed and expected SM background events in the two SRs  
are summarised in Table~\ref{table:2lepton_expect_observed}, for $ee$ and 
$\mu\mu$ channels summed. The $ee$ and $\mu\mu$ contributions are also 
shown separately for illustration. In all SRs, the SM expectation and 
observation agree within uncertainties. 

\begin{figure}[t!]
  \begin{center}
    \includegraphics[width=8cm]{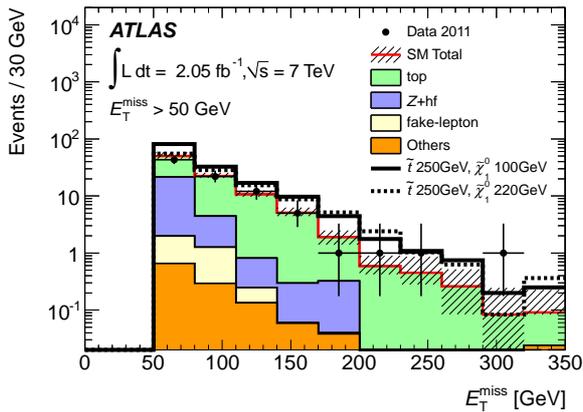}
  \end{center}
  \caption{
  The $ee$+$\mu\mu$ $\etmiss$ distribution for SR1 compared to the SM predictions, shown by the light (red) solid line, 
  and SM+signal predictions, shown by the dark (black) solid and dashed lines. 
  The dashed band represents the total systematic uncertainty. The last \MET bin includes the number of overflow
  events for both data and SM expectation.}
  \label{figure:2lepton_signalregionplots_sum}
\end{figure}

In Figure~\ref{figure:2lepton_signalregionplots_sum} the distributions 
of $E_{\rm T}^{\rm miss}$ in SR1 (full spectrum) and SR2 ($E_{\rm T}^{\rm miss}>$80~GeV), 
summing the $ee$ and $\mu\mu$ channels, are shown. For illustrative purposes, the 
distributions expected for two signal ($\tone$,$\neut$) scenarios with masses of 
(250,100)~GeV and (250,220)~GeV, respectively, are added to the SM predictions.

\begin{table}[t!]
  \begin{center}
    \begin{tabular}{ccc}
      \hline
      & SR1 & SR2  \\
      \hline
      \multicolumn{3}{c}{$ee$ channel} \\
      \hline
      Data  & 39 & 20  \\
      SM & 36.2$\pm$8.5 & 14.1$\pm$3.0 \\
      \hline
      top & 23.8$\pm$4.8 & 11.9$\pm$2.8  \\
      $Z+$hf & 9.4$\pm$7.0 & 0.9$\pm$0.8  \\
      fake lepton & 2.4$\pm$0.9  & 1.1$\pm$0.6 \\
      Others & $0.5\pm 0.5$ & $0.2\pm 0.2$ \\
      \hline
      \hline
      \multicolumn{3}{c}{$\mu\mu$ channel} \\
      \hline
      Data & 47  & 23  \\
      SM & $55\pm 12$  & 26.6$\pm$5.1 \\
      \hline
      top & $40.4\pm 6.2$ & $22.9\pm 4.3$  \\
      $Z+$hf & 14.2$\pm$9.9 & 3.3$\pm$2.6  \\
      fake lepton & 0.00$\pm$0.08 & 0.00$\pm$0.07  \\
      Others & $0.7\pm 0.7$ & $0.3\pm 0.3$ \\
      \hline
      \hline
      \multicolumn{3}{c}{$ee$+$\mu\mu$} \\
      \hline
      Data  & 86 & 43 \\
       SM & 92$\pm$19 & $40.7\pm 6.0$ \\
      \hline
      top & 64.3$\pm$7.7 & $34.8\pm 5.0$ \\
      $Z+$hf & 24$\pm$16 & 4.2$\pm$3.2 \\
      fake lepton & $2.4\pm 0.9$ & $1.1\pm 0.6$ \\
      Others & $1.2\pm 1.2$ & $0.6\pm 0.6$ \\
      \hline
      \multicolumn{3}{c}{ } \\
      \multicolumn{3}{c}{95\% C.L. upper limits: observed (expected)}  \\
      \hline
      events  & 37.2 (40.6) & 19.8 (17.8) \\
      visible $\sigma$ [fb] & 18.2 (19.8) & 9.7 (8.7) \\
      \hline
    \end{tabular}
  \end{center}
  \caption{Expected and measured number of events in SR1 and SR2 for $ee$ and $\mu\mu$ channels 
  (separately and summed) for an integrated luminosity of \lumi~fb$^{-1}$.
  Rows labelled as `Others' correspond to the subdominant SM backgrounds estimated from MC simulation.
  The total systematic uncertainties are also displayed. 
  At the bottom, model-independent observed and expected limits at 95\% C.L on the number of events and 
  visible cross sections are shown summing the $ee$ and $\mu\mu$ channels. }
  \label{table:2lepton_expect_observed}
\end{table}

\begin{figure}[t!]
  \begin{center}
    \includegraphics[width=8cm]{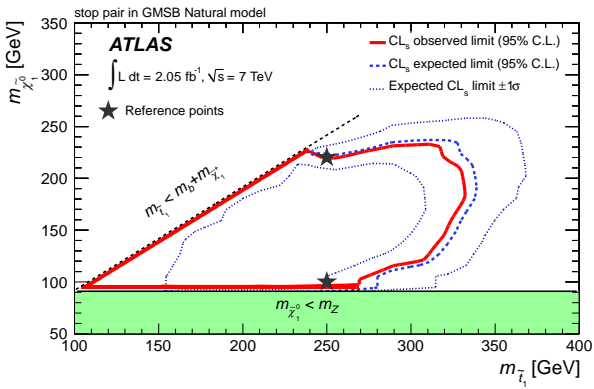}
  \end{center}
  \caption{
  Expected and observed exclusion limits and $\pm 1 \sigma$ variation on the expected limit 
  in the $\tone$-$\neut$ mass plane. The reference points indicated on the plane correspond to 
  the ($\tone$,$\neut$) scenarios of (250,100)~\GeV and (250,220)~\GeV, respectively.}
  \label{figure:stopGMSB_exclusionlimit}
\end{figure}

The results are translated into 95\% confidence level (C.L.) upper limits on 
contributions from new physics using the CL$^{}_{\rm s}$ prescription~\cite{Read:2002}. 
The SR with the better expected sensitivity at each point in parameter space 
is adopted as the nominal result. Systematic uncertainties are treated as nuisance 
parameters and their correlations are taken into account. 
Figure~\ref{figure:stopGMSB_exclusionlimit} shows the observed and expected exclusion 
limits at 95\% C.L. in the $\tone$-$\neut$ mass plane, assuming direct stop pair production 
in the framework of GMSB models with light higgsinos. The $\pm 1\sigma$ contours around 
the median expected limit are also shown. 
Stop masses up to 310~\GeV\, are excluded for 115~\GeV $<m_{\tilde{\chi}_{1}^{0}}<230$~\GeV. 
The exclusion extends to stop masses of 330 \GeV\, for a neutralino mass of about 190~\GeV.  
Stop masses below $240$ \GeV\, are excluded for $m_{\tilde{\chi}_{1}^{0}}>m_{Z}$.
The two SRs are used to set limits on the number of events and the visible cross section, $\sigma^{}_{\rm vis}$,  
of new physics models, without corrections for the effects of experimental resolution, acceptance and efficiency. 
The observed and expected excluded values at 95\% C.L. are reported 
in Table~\ref{table:2lepton_expect_observed}. 

\section{Conclusions} 
In summary, results of a search for direct scalar top quark pair production in $pp$ 
collisions at $\sqrt{s}$ = 7~TeV, based on 2.05~fb$^{-1}$ of ATLAS data are reported. 
Scalar top quarks are searched for in events with two same flavour opposite-sign 
leptons ($e,\mu$) with invariant mass consistent with the $Z$ boson mass, 
large missing transverse momentum and jets in the final state, 
where at least one of the jets is identified as originating from a {\it b-}quark. 
The results are in agreement with the SM prediction and are interpreted 
in the framework of $R$-parity conserving `natural' gauge mediated SUSY 
scenarios. Stop masses up to 310~\GeV\, are excluded for 115~\GeV~$< m_{\tilde{\chi}_{1}^{0}}<$~230~\GeV\, at 
95\% C.L., reaching an exclusion of $m_{\tone}<$~330~\GeV\, for $m_{\tilde{\chi}_{1}^{0}}=190$~\GeV. 
Stop masses below $240$~\GeV\, are excluded for $m_{\tilde{\chi}_{1}^{0}}>m_{Z}$.

\section{Acknowledgements}

We thank CERN for the very successful operation of the LHC, as well as the
support staff from our institutions without whom ATLAS could not be
operated efficiently.

We acknowledge the support of ANPCyT, Argentina; YerPhI, Armenia; ARC,
Australia; BMWF, Austria; ANAS, Azerbaijan; SSTC, Belarus; CNPq and FAPESP,
Brazil; NSERC, NRC and CFI, Canada; CERN; CONICYT, Chile; CAS, MOST and
NSFC, China; COLCIENCIAS, Colombia; MSMT CR, MPO CR and VSC CR, Czech
Republic; DNRF, DNSRC and Lundbeck Foundation, Denmark; ARTEMIS, European
Union; IN2P3-CNRS, CEA-DSM/IRFU, France; GNAS, Georgia; BMBF, DFG, HGF, MPG
and AvH Foundation, Germany; GSRT, Greece; ISF, MINERVA, GIF, DIP and
Benoziyo Center, Israel; INFN, Italy; MEXT and JSPS, Japan; CNRST, Morocco;
FOM and NWO, Netherlands; RCN, Norway; MNiSW, Poland; GRICES and FCT,
Portugal; MERYS (MECTS), Romania; MES of Russia and ROSATOM, Russian
Federation; JINR; MSTD, Serbia; MSSR, Slovakia; ARRS and MVZT, Slovenia;
DST/NRF, South Africa; MICINN, Spain; SRC and Wallenberg Foundation,
Sweden; SER, SNSF and Cantons of Bern and Geneva, Switzerland; NSC, Taiwan;
TAEK, Turkey; STFC, the Royal Society and Leverhulme Trust, United Kingdom;
DOE and NSF, United States of America.

The crucial computing support from all WLCG partners is acknowledged
gratefully, in particular from CERN and the ATLAS Tier-1 facilities at
TRIUMF (Canada), NDGF (Denmark, Norway, Sweden), CC-IN2P3 (France),
KIT/GridKA (Germany), INFN-CNAF (Italy), NL-T1 (Netherlands), PIC (Spain),
ASGC (Taiwan), RAL (UK) and BNL (USA) and in the Tier-2 facilities
worldwide.

\bibliography{PUB_GMSB}

\clearpage 
\onecolumn
\begin{flushleft}
{\Large The ATLAS Collaboration}

\bigskip

G.~Aad$^{\rm 48}$,
B.~Abbott$^{\rm 111}$,
J.~Abdallah$^{\rm 11}$,
S.~Abdel~Khalek$^{\rm 115}$,
A.A.~Abdelalim$^{\rm 49}$,
O.~Abdinov$^{\rm 10}$,
B.~Abi$^{\rm 112}$,
M.~Abolins$^{\rm 88}$,
O.S.~AbouZeid$^{\rm 158}$,
H.~Abramowicz$^{\rm 153}$
% ...
\bigskip

% ...
$^{10}$ Institute of Physics, Azerbaijan Academy of Sciences, Baku, Azerbaijan\\
$^{11}$ Institut de F\'isica d'Altes Energies and Departament de F\'isica de la Universitat Aut\`onoma  de Barcelona and ICREA, Barcelona, Spain\\
% ...
$^{48}$ Fakult\"{a}t f\"{u}r Mathematik und Physik, Albert-Ludwigs-Universit\"{a}t, Freiburg i.Br., Germany\\
$^{49}$ Section de Physique, Universit\'e de Gen\`eve, Geneva, Switzerland\\
% ...
$^{88}$ Department of Physics and Astronomy, Michigan State University, East Lansing MI, United States of America\\
% ...
$^{111}$ Homer L. Dodge Department of Physics and Astronomy, University of Oklahoma, Norman OK, United States of America\\
$^{112}$ Department of Physics, Oklahoma State University, Stillwater OK, United States of America\\
% ...
$^{115}$ LAL, Univ. Paris-Sud and CNRS/IN2P3, Orsay, France\\
% ...
$^{153}$ Raymond and Beverly Sackler School of Physics and Astronomy, Tel Aviv University, Tel Aviv, Israel\\
% ...
$^{158}$ Department of Physics, University of Toronto, Toronto ON, Canada\\
% ...
\end{flushleft}


\end{document}